\documentclass[conference]{IEEEtran}
\IEEEoverridecommandlockouts
\usepackage{cite}
\usepackage{amsmath,amssymb,amsfonts}
\usepackage{algorithmic}
\usepackage{graphicx}
\usepackage{textcomp}
\usepackage{amsmath}
\usepackage[ruled]{algorithm2e}
\usepackage{graphicx}
\usepackage{textcomp}
\usepackage{xcolor}
\usepackage{balance}
\usepackage{booktabs}
\usepackage{listings}
\colorlet{punct}{red!60!black}
\definecolor{background}{HTML}{EEEEEE}
\definecolor{delim}{RGB}{20,105,176}
\colorlet{numb}{magenta!60!black}
\lstdefinelanguage{json}{
    basicstyle=\small,
    numberstyle=\scriptsize,
    stepnumber=1,
    numbersep=8pt,
    showstringspaces=false,
    breaklines=true,
    frame=lines,
    backgroundcolor=\color{background},
    literate=
     *{0}{{{\color{numb}0}}}{1}
      {1}{{{\color{numb}1}}}{1}
      {2}{{{\color{numb}2}}}{1}
      {3}{{{\color{numb}3}}}{1}
      {4}{{{\color{numb}4}}}{1}
      {5}{{{\color{numb}5}}}{1}
      {6}{{{\color{numb}6}}}{1}
      {7}{{{\color{numb}7}}}{1}
      {8}{{{\color{numb}8}}}{1}
      {9}{{{\color{numb}9}}}{1}
      {:}{{{\color{punct}{:}}}}{1}
      {,}{{{\color{punct}{,}}}}{1}
      {\{}{{{\color{delim}{\{}}}}{1}
      {\}}{{{\color{delim}{\}}}}}{1}
      {[}{{{\color{delim}{[}}}}{1}
      {]}{{{\color{delim}{]}}}}{1},
}
\usepackage{xcolor}
\def\BibTeX{{\rm B\kern-.05em{\sc i\kern-.025em b}\kern-.08em
    T\kern-.1667em\lower.7ex\hbox{E}\kern-.125emX}}
\begin{document}

\title{Fostering Trust in Smart Inverters: A Framework for Firmware Update Management and Tracking in VPP Context}

\author{
    \IEEEauthorblockN{Thusitha Dayaratne\IEEEauthorrefmark{1}, Carsten Rudolph\IEEEauthorrefmark{1}, Tom Shirley\IEEEauthorrefmark{2}, Sol Levi\IEEEauthorrefmark{2}, David Shirley\IEEEauthorrefmark{2}
    }
    \IEEEauthorblockA{\IEEEauthorrefmark{1}Faculty of IT, Monash University, Melbourne, Australia
   }
    \IEEEauthorblockA{\IEEEauthorrefmark{2}Selectronic, Melbourne, Australia\\\{thusitha.dayaratne, carsten.rudolph\}@monash.edu
    \\\{tshirley, slevi, dshirley\}@selectronic.com.au}
     
}
\maketitle

\begin{abstract}
Ensuring the reliability and security of smart inverters that provide the interface between distributed energy resources (DERs) and the power grid becomes paramount with the surge in integrating DERs into the (smart) power grid. Despite the importance of having updated firmware / software versions within a reasonable time frame, existing methods for establishing trust through firmware updates lack effective historical tracking and verification. This paper introduces a novel framework to manage and track firmware update history, leveraging verifiable credentials. By tracking the update history and implementing a trust cycle based on these verifiable updates, we aim to improve grid resilience, enhance cybersecurity, and increase transparency for stakeholders.

\end{abstract}

\begin{IEEEkeywords}
Smart inverters, Verifiable credentials, Secure firmware updates, Virtual power plants
\end{IEEEkeywords}

\section{Introduction}
Solar photovoltaic (PV) is considered the fastest-growing form of energy generation, claiming approximately 10\% and 5\% of the electricity generated in Australia and Europe, respectively. In particular, solar PV accounts for combined capacities of more than 11 GW in Australia and 18 GW in Europe~\cite{solarenergyAu,solarenergyEU}. This growing penetration of distributed energy resources (DERs) provides more opportunities for greener energy. To effectively manage distributed generation, distributed system operators (DSOs) leverage virtual power plants (VPPs) by combining the DERs and the willingness of consumers to share their DERs with financial benefit. A VPP is a logical grouping of a set of DERs in a defined geographical area as a single-generation unit~\cite{saboori2011virtual} managed by an independent entity such as a VPP operator. 

VPPs rely on smart inverters, as inverters are the primary interface between consumers/utilities and DERs (both for batteries and solar panels). Consumers possess legal contracts with the VPP operator, which define the VPP operator's access to the inverter/DER and control over how these are integrated into the VPP and define financial incentives for consumers. Further, the control and monitoring capabilities of the VPP operator on inverters are restricted and mandated by the legal and regulatory context of the environment. For example, utilities should be able to stop the output of consumer inverters to the energy grid in Australia~\cite{aemoinverters2020}. The flexibility of VPPs comes with a price, as wide-area communication and remote control/sensing between smart inverters and utilities/operators significantly expand the attack surface for cyberattacks~\cite{qi2016cybersecurity}. This risk increases further, given that smart inverters/DERs reside outside the network operator's control perimeter. 

Security approaches, including network segmentation~\cite{johnson2017design}, encryption, and zero-trust~\cite{Alagappan2022} are used to thwart attacks that exploit vulnerabilities in commonly used standards such as SunSpec Modbus and DNP3. Nevertheless, adversaries can bypass these security measures by leveraging software-based attacks in compromising firmware/software of equipment to gain access~\cite{zhang2018security,Choi2021,konstantinou2015impact}. Thus, having a reliable firmware update process is essential to ensure the security of VPP functionalities. Nevertheless, the current firmware version information of an inverter alone is not adequate, given that the adversaries can mimic to possess the latest firmware version if they compromise the inverter while possessing an earlier vulnerable firmware version. Hence, the assurance that the inverter consistently runs firmware versions without known security vulnerabilities and is updated within reasonable time windows by assessing the complete update history is of utmost importance.

In this work, we employ the concept of verifiable credentials (VCs) to enhance the security of the inverter firmware/software update process by tracking the firmware history. Additionally, we introduce a novel framework designed to enhance the security of VPP operations by effectively managing interactions between inverters and VPP operators based on firmware updates. Specifically, we investigate the possibility of providing security assurance without resorting to remote attestation (RA) or a hardware security module such as a trusted platform module. Our contributions are outlined as follows:
\begin{enumerate}
    \item Present a novel method leveraging verifiable credentials for establishing a comprehensive update history tracking system complemented by VC based firmware update mechanism 
    \item Introduce a trust cycle leveraging the firmware update history to establish secure interactions between inverters and VPP operators, enhancing security in the integration process
    \item Assess the proposed scheme in comparison to the well-established IEEE 2030.5-2018/CSIP standard/protocol and proof-of-concept implementation highlighting its distinctive features, potential advancements and performance
\end{enumerate}

\section{Related Work}
Given the novelty of the decentralised VPP concept, only a few works have analysed reporting on inverter firmware updates. Bere et al.~\cite{Bere2021} used blockchain to distribute keys and smart contracts to verify the authenticity and integrity of the firmware in proposing a framework for firmware security checks and recovery for smart inverters. Ahn et al.~\cite{Ahn2021} presented a proof of concept for the proposed model as an onboard security module that transforms traditional inverters into smart inverters to support firmware updates. Choi et al.~\cite{Choi2021} further extended this by integrating a physically unclonable function (PUF) into the onboard security model, improving the security of cryptography operations. Ansay et al.~\cite{Ansay2019} used decentralised identifiers and VCs in designing a framework that supports software updates for 5G-enabled IoT devices. Nevertheless, to the best of our knowledge, no work has looked into leveraging VC to manage and track the firmware updates. Moreover, no prior literature uses the history of updates to support VPP operators in the risk estimation and to determine the level of interaction in the context of the smart grid. 

\section{Use case}
\begin{figure*}
    \centering
    \includegraphics[width=\linewidth]{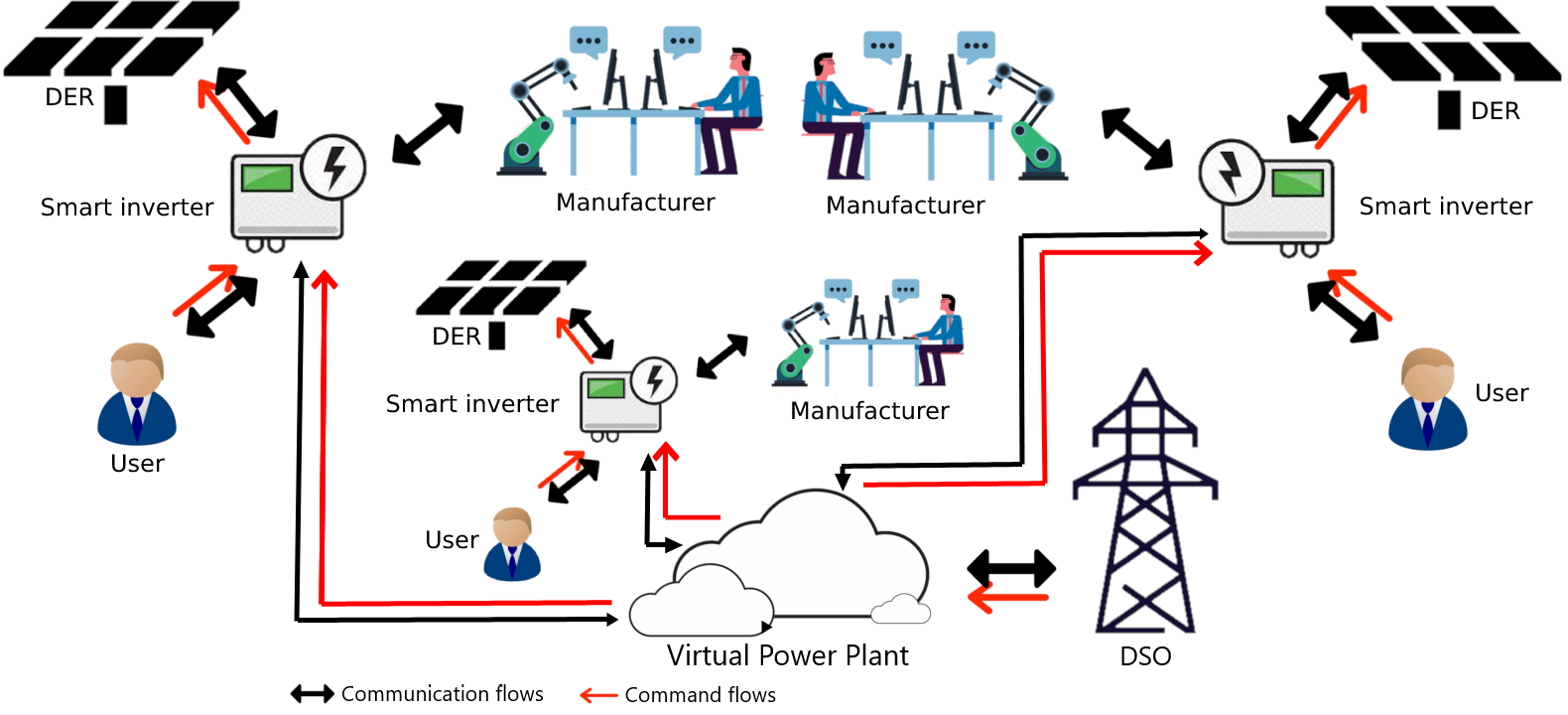}
    \caption{Entities and their interaction flows}
    \label{fig:usecase}
\end{figure*}

In this work, we analyse the use of smart inverters within a microgrid scenario, where these intelligent devices play a pivotal role in ensuring demand-supply balance as part of an aggregator/VPP setup. The interaction between the grid entities is shown in Figure~\ref{fig:usecase}. Individuals who own DERs, such as solar panels and battery storage units, procure smart inverters from manufacturers/sellers and integrate them into their setups. Each DER is connected to a smart inverter that manages, monitors, and regulates the associated DER. Although practical setups often include gateways and home energy management systems (HEMS), we simplify this by treating the combination of inverter and gateway/HEMS as a single entity. Every smart inverter is assumed to possess a dependable static identifier assigned during the manufacturing phase. These inverters are equipped with firmware to control the underlying hardware and software to facilitate interaction between the inverter and the user/operator. Manufacturers release updated firmware/software versions to rectify bugs, address security vulnerabilities, and enhance performance as needed. Users can choose manual or automatic updates for their inverter firmware.

We assume that a group of consumers enrol their inverters and DERs into a VPP. Upon enrolment, the VPP operator sends control or monitoring signals to all enrolled inverters or a designated subset. These inverters authenticate the legitimacy of the commands and execute them. Subsequently, the requested data or the resulting status information is sent to the VPP operator. The operator then validates these responses, assessing the condition of each device in relation to the broader grid and the VPP condition.

\subsection{Security Requirement}
This section presents the basic security requirements of the proposed system. 

\subsubsection{Confidentiality}
Confidentiality ensures that sensitive information remains private and is not disclosed to unauthorised entities. In the considered context, confidentiality is essential to restrict access to specific information associated with a particular inverter, such as its firmware version and update date-time. 

\subsubsection{Integrity}
Integrity guarantees that data or information remains unchanged and unaltered during transmission or processing. In the VPP context, integrity ensures that firmware updates released by manufacturers reach smart inverters without unauthorised alterations. Any modification to the firmware during transit could compromise security or functionality. Additionally, integrity ensures that the information received by an entity about a specific inverter is the same information sent, free from tampering or alterations.

\subsubsection{Availability}
Availability ensures that a system or service is operational and accessible when needed. In our system, availability ensures that authorised entities such as DNSPs and VPP operators can access information about specific inverters promptly. This uninterrupted access is vital for efficient grid management and VPP operations.

\subsubsection{Authentication}
Authentication is the process of verifying the identity of a user, device, or entity to confirm their claimed identity. In the described context, authentication ensures that manufacturers and users are who they claim to be and that smart inverters are authentic and have not been tampered with. It thwarts unauthorised firmware updates and ensures that only legitimate inverters participate in VPP operations.

\subsubsection{Authorisation}
Authorisation involves granting or denying access to specific resources or actions based on authenticated identity and permissions. In our system, authorisation dictates who can install new updates for inverters, request verification of inverter information, and validate information about installed firmware versions. Having a proper authorisation ensures the safe operation of the VPP.

\subsubsection{Non-Repudiation}
Non-repudiation is the assurance that an entity involved in a communication or transaction cannot deny its participation or the actions it performed. In the described scenario, non-repudiation ensures that when manufacturers release firmware/software updates for smart inverters, the manufacturer cannot deny releasing the update, and the inverters receiving and installing these updates cannot later deny having applied them. This feature ensures accountability and trust in update processes.

\section{Threat Model}

This work only focusses on the threat of inverters that run firmware with known vulnerabilities or already compromised firmware versions. Adversaries can potentially execute several attacks, given that an inverter is not up-to-date with the latest secure firmware version or running a compromised version. Two potential attacks that arise from the threat and are connected to the system model are described below. 

\vspace{-0.1cm}
\subsection{Attack 1: Mimic incorrect meta information of an inverter}
Adversaries can mimic information about inverter capabilities. For example, an adversary can attempt to mimic that their inverter possesses the latest firmware version. Similarly, an adversary may over- or under-represent inverter capabilities, e.g. min-max limits, mode, etc. Successful attempts can result in incorrect/sub-optimal power flow calculations for the connected community, diminishing the potential advantages of the VPP.

\vspace{-0.1cm}
\subsection{Attack 2: Enrol unauthorised inverters}
A dishonest user/intruder can enrol an inverter that does not match the enrolment criteria of the particular VPP program, given the compromised firmware version. This type of attack is plausible as the adversary can mimic incorrect information. Further, adversaries may enrol an inverter that is not owned by them. Successful enrolment of an unauthorised inverter can have varying impacts on the community and the legitimate inverter owner.

\vspace{-0.5cm}
\section{Preliminaries}
\subsection{Decentralised Identifiers (DIDs)}
A DID is a globally unique and persistent identifier. DIDs enable verifiable and decentralised digital identities for entities \cite{didw3c} that are under total control of the identity owner, instead of relying on a central authority. The W3C (World Wide Web Consortium) approved the DID specification as a recommendation in late June 2022. The DID representation conforms with the URI (universal resource identifier) scheme with three mandatory components as below.
\[
\color{blue}did:\color{magenta}<did\textnormal{-}method>:\color{teal}<method\textnormal{-}specific\textnormal{-}identifier>
\]

\begin{figure}
    \centering
    \includegraphics[width=\linewidth]{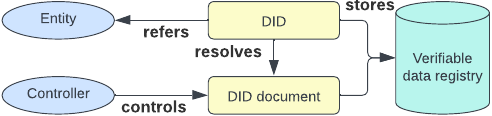}
    \caption{High level view of the DID architecture}
    \label{fig:did_architecture}
\end{figure}

A DID resolves to a DID document (DIDdoc), which is a JSON-LD (JavaScript Object Notation-Linked Data) object. The DIDdoc possesses public keys for cryptographic operations, including authentication and verification. The DIDdoc can also contain service endpoints, which can be used to advertise different services provided by the entity or the controller of the entity. Figure \ref{fig:did_architecture} depicts the overview of the DID architecture. 

\subsection{Verifiable Credentials (VCs)}
The W3C introduced the concept of VCs, which are portable and provable claims about an entity. VCs represent statements made by an issuer in a tamper-evident and privacy-respecting manner \cite{vcw3c}. In contrast to digital certificates, VCs provide ultimate flexibility in the stricture and distributed trust eliminating the single point of failure. A VC consists of three components: metadata, claims, and proof. Metadata describes meta-information about the VC, such as the issuer, expiry date, and public keys for verification. Claims depict a set of attributes about the entity and the proof is to provide tamper-evident. Similar to the DIDs, VCs also leverage verifiable data registries (VDRs). 

\begin{figure}
    \centering
    \includegraphics[width=\linewidth]{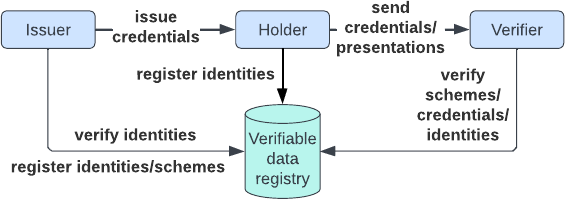}
    \caption{High level architecture of the verifiable credentials}
    \label{fig:vc_architecture}
\end{figure}

The VC ecosystem is illustrated in Figure \ref{fig:vc_architecture}, encompassing the supporting VDR and the three main entities: issuer, holder, and verifier. The credibility of a specific VC depends on the reputation of the issuer. Consequently, VCs are typically issued by reputable or trustworthy entities in practical scenarios, although any entity can act as a VC issuer. Holders securely store their VCs in a digital wallet and present them to verifiers upon request. Verifiers can verify the presented VCs without direct communication with the issuer, as corresponding keys can be retrieved from the VDR. This approach prevents issuers from establishing connections between the holder and the services they access.

\section{Proposed Model - VCs for Secure Inverter Management in VPP}
We explain the proposed model and the assumed setup in this section. All the notation used throughout the paper is shown in Table \ref{tab:Notations}.

\begin{table}
\caption{\label{tab:Notations}Notations}
\centering
\begin{tabular}{p{1in}p{2in}}
\toprule
\textbf{Symbol} & \textbf{Definition}\tabularnewline
\midrule 
M & Manufacturer \tabularnewline
SC & Smart Contract\tabularnewline
SI & Smart Inverter\tabularnewline
U & Inverter owner\tabularnewline
$FW_{X}^i$ & Firmware version \textit{i} issued by \textit{X}\tabularnewline
$VC_{X}^{Y}$ & VC of \textit{Y} issued by \textit{X}\tabularnewline
$DID_{X}^P$ & Public DID of entity \textit{X}\tabularnewline
$SK_{X}^{DID}$ & Private key associated with \textit{X}'s public DID\tabularnewline
$PK_{X}^{DID}$ & Public key associated with \textit{X}'s public DID\tabularnewline
$DIDDoc_{X}^P$ & DIDdoc of \textit{X}'s public DID\tabularnewline
$hash(X)$ & Hash function\tabularnewline
$sign(data, key)$ & Digital signature function\tabularnewline
$encrypt(data, key)$ & Encryption function\tabularnewline
$nonce$ & Nonce\tabularnewline
$||$ & Concatenation operation\tabularnewline
\bottomrule 
\end{tabular}
\end{table}

\subsection{Setup}
The general public does not participate in the use case. Therefore, the use of a public blockchain is unnecessary. In contrast, a private blockchain is unsuitable given the multiple stakeholders. Thus, we propose using a consortium blockchain in the design, as it provides the correct amount of security and decentralisation. We assume that each user has a key pair to interact with the chain. There exists a DID API to manage DIDs. This API provides functionalities including generating DIDs, updating DIDdocs, and retrieving existing DIDdocs (DID resolving). Additionally, we assume the following throughout this work, which are explained in detail in appropriate sections.  

\begin{enumerate}
    \item Inverter manufacturers possess public DIDs
    \item Each smart inverter is assigned a public DID by its manufacturer
    \item Smart inverters possess a digital wallet to store keys and VCs
    \item Manufacturers register DIDs and VC schemes on a suitable blockchain platform
    \item There is a secure communication channel between the inverter and manufacturer
    \item Manufacturers possess a mechanism to determine the firmware version of an inverter and its installed datetime
    \item Successful bootstrapping of inverters
\end{enumerate}

All these assumptions can be realised with software/firmware solutions. Thus, in principle, there is no significant cost increase for manufacturers as no additional hardware components are required. However, higher levels of security can potentially be achieved by integrating secure hardware components such as trusted platform modules.

\subsection{Part 1: Tracking Inverter Firmware Updates}
\begin{figure}
    \centering
    \includegraphics[width=\linewidth]{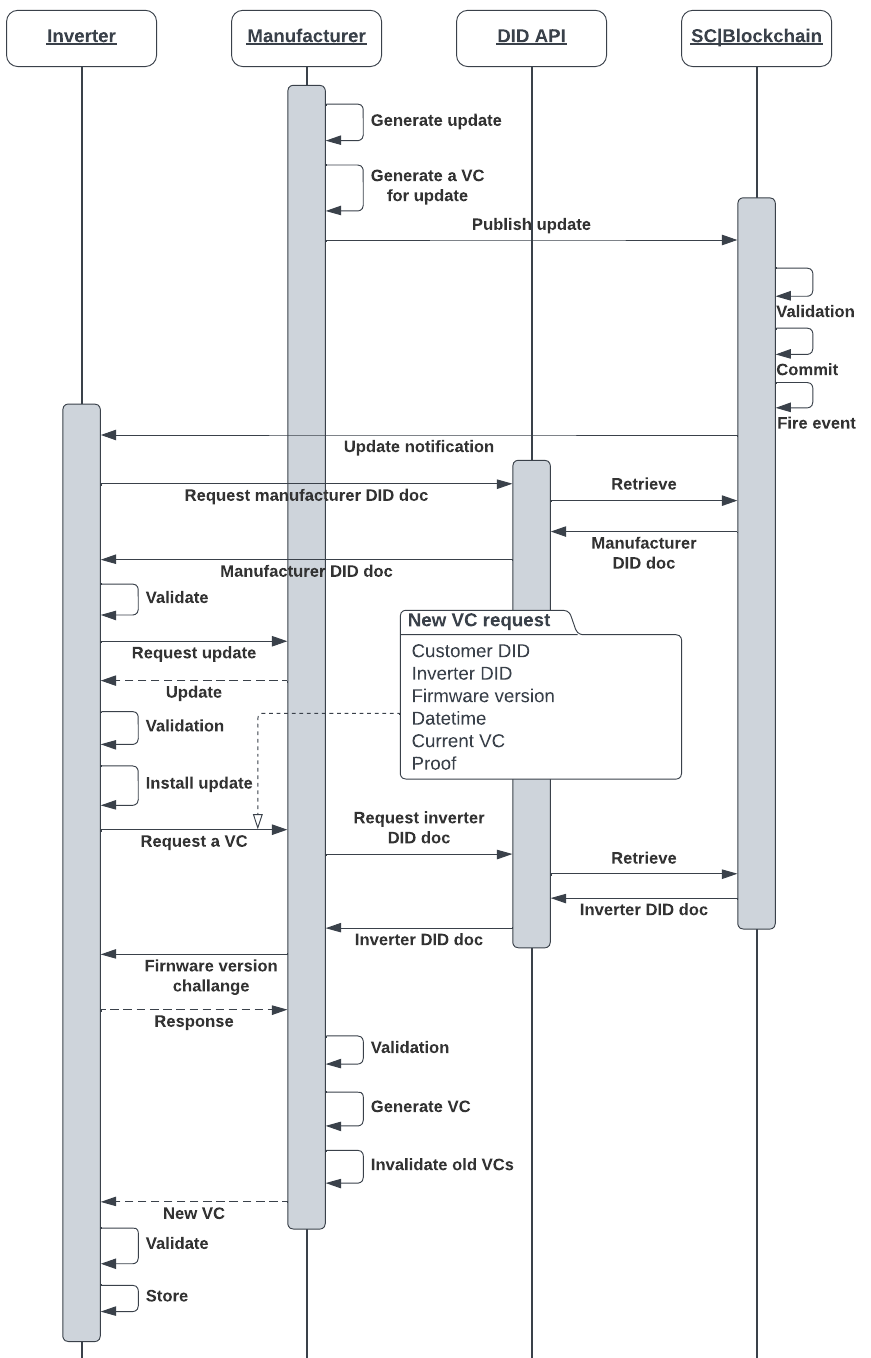}
    \caption{Sequence diagram for the firmware update process}
    \label{fig:seq_update_process}
\end{figure}

This section introduced the VC-based security model that improves the security of VPP operations by tracking inverter firmware updates. The entire process from update manufacturing to reporting firmware updates, is depicted in Figure \ref{fig:seq_update_process}. We divide the process into multiple sub-processes for brevity. 

\subsubsection{Firmware Update Management Smart Contract}
The manufacturer deploys a smart contract into the blockchain to manage the availability of its firmware updates. This smart contract emits events whenever a new update is made available. The smart contract should incorporate suitable validation mechanisms to ensure that only the manufacturer has the authorisation to introduce new update information. This measure prevents malicious users within the system from injecting false update data into the ecosystem. A sample smart contract is shown in Listing \ref{lst:sc_update}.

\begin{lstlisting}[caption = Sample smart contract,captionpos=b,label=lst:sc_update]
|\color{blue}contract| ManufacturerUpdate {

  |\color{blue}address| |\color{teal}private| manufacturer;

  |\color{magenta}constructor()| {
    manufacturer = |\color{blue}msg|.sender;
  }
  
  |\color{blue}event| NewUpdate(|\color{blue}address| |\color{yellow}indexed| from, |\color{blue}string| version, |\color{blue}string| model, |\color{blue}string| credential);

  |\color{blue}function| SaveUpdate(|\color{blue}address| _to, |\color{blue}string| _version, |\color{blue}string| _model, |\color{blue}string| _credential) |\color{teal}public| {
    |\color{cyan}require|(manufacturer == |\color{cyan}msg|.sender, |\color{pink}"Only manufacturer can push updates"|);
    |\color{gray}// Save update VC logic here|
    |\color{blue}emit| NewUpdate(|\color{cyan}msg|.sender, _version, _model, _credential);
  }
}
\end{lstlisting}

\subsubsection{Inverter Manufacturing}
The manufacturer generates a key pair (a public and a private key) and assigns a public DID generated using the DID API for every inverter. The corresponding DIDdoc is stored on a blockchain or a distributed file system, such as IPFS, based on the DID API implementation. Once the DIDdoc is published, the manufacturer stores the inverter's key pair, the DID, and the manufacturer's DID in inbuilt inverter wallet. Furthermore, the manufacturer subscribes the inverter to the update smart contract's events. Subscribing and listening functionalities can be integrated into the inverter's software by leveraging established Web3 providers and libraries, such as MetaMask and Web3.js/ether.js. 

The owner must possess the corresponding private key to manage the inverter DIDdoc. Thus, the manufacturer must either securely pass the associated key pair of the inverter to the inverter owner or update the DIDdoc to reflect the owner-provided public key upon a sale. Additionally, it should be noted that the proposed model does not require changes to any existing update delivery methods. However, for completeness and clarity, we elaborate on how to leverage the VC to publish new updates from the manufacturer's perspective and how to install updates from the inverter's perspective in the next two subsections.

\subsubsection{Publishing New Updates}
The manufacturer ($M$) is responsible for generating the firmware update ($FW_{M_i}^i$) along with the corresponding VC ($VC_{M_i}^{FW_i}$) for the update. The proposed structure of this VC is depicted in Listing \ref{lst:vc_update}. This VC encapsulates vital details concerning the update. Specifically, we suggest incorporating the manufacturer's DID, release date, downloadable link, version number, hash value of the binary, type of the update (e.g. security, bug, feature, etc.) and a list of supported inverter models within the VC. Additionally, we propose including the CVE (Common Vulnerabilities and Exposures) numbers that are addressed from the update if it is a security update. 

Updates can be stored directly on the blockchain or within an update server. Upon generation of the VC, the constructed VC and the public DID of the manufacturer ($DID_{M_i}^P$) are sent to the smart contract ($SC$). The manufacturer's private key (corresponding to the DID) ($SK_{M_i}^{DID}$) is used to sign the message to ensure integrity and authenticity. The smart contract verifies the message signature using the manufacturer's public key ($PK_{M_i}^{DID}$) specified in the corresponding DIDdoc ($DIDDoc_{M_i}^P$). Once the verification is a success, the VC is committed to the blockchain, which triggers a smart contract event to notify the availability of the new firmware version.

\begin{lstlisting}[caption = Sample VC for an update,captionpos=b,label=lst:vc_update,language=json]
{
  "@context":[..],
  "id": "did:sov:sg:inverter:firmware:vc:123456789",
  "type": ["VerifiableCredential","FirmwareVC"],
  "issuer": "did:sov:sg:manufacturer:123456789",
  "issuanceDate": "2023-04-01T10:11:12Z",
  "credentialSubject": {
    "manufacturer": "did:sov:sg:manufacturer:123456789",
    "releasedDate": "2023-04-01T10:00:00Z",
    "link": "<url to download the binary>",
    "frimwareInfo": {
        "version": "1.0021",
        "binaryHash": "ARVDVX2753hd6752H",
        "type": "security"
    },
    "associatedCVEs": {
        |\color{blue} //list of CVE numberd that are addressed from the update|
    },
    "supportingModels":{
        |\color{blue} //all the model numbers that supports the firmware|
    }
  },
  ...
  "proof": {..}
}
\end{lstlisting}

\subsubsection{Firmware Update Process}
Once the inverter is installed in a home, the firmware can be updated whenever a newer version is available. 
Inverters ($SI$) get notified of the availability of new updates, as inverters are listening to the events of the corresponding manufacturer's smart contract. Upon receiving an event, the inverter checks if the model matches and if the version is newer than the current version of the inverter. If both conditions are satisfied, the inverter saves the received VC for the update in its wallet. The inverter can directly refer to the download link specified in the VC to download the update. In scenarios where the link does not exist on the VC, the inverter requests the manufacturer's DIDdoc from the DID API using the embedded manufacturer's DID. The DIDdoc lists the manufacturer's update endpoint. Inverter extracts the endpoint from the document and requests the update from the update server. Upon receiving the update, the inverter verifies the integrity of the update by comparing the received update hash ($H_{FW_i}^{received}$) with the update hash ($binaryHash$) stated in the VC received from the smart contract. The inverter installs the update if the verification succeeds or discards the update if it fails. 
We emphasise the assumption of secure bootstrapping of the inverter, as it is pivotal in establishing the root-of-trust for the firmware update process. 

\subsubsection{Issue a New VC for the Inverter}
A successful update updates the firmware version of the inverter. Nevertheless, the newer version will not automatically reflect on the firmware version stated in the inverter's existing VC. Thus, a new VC should be generated, and the current VC should be invalidated to reflect the changes in the update history. 

The user requests a new VC from the manufacturer using the VC endpoint listed in the manufacturer's DIDdoc. The request consists of the customer DID, inverter DID, firmware version, installed date-time, current VC, and a nonce. Hardware-based RA is the best solution to obtain the installation date, time, and version of the installed firmware. However, given the unavailability of hardware-based RA methods, we assume that the manufacturer embeds a software-based method (software-based RA) to determine the firmware version of an inverter and its installed date-time. A potential method to implement the update-specific string is the use of an update-specific, one-time password approach, similar to the time-based one-time password \cite{totp} method. Users use the update-specific string ($P_i$) generated from the update instead of manually specifying the update version and date-time. The request should be signed using the user's private key ($SK_{U_i}^{DID}$) corresponding to the public DID $PK_{U_i}^{DID}$ of the user. The manufacturer's public key can be used to encrypt the signed message to ensure the integrity and privacy of inverter details.

\begin{lstlisting}[caption = Sample VC for an inverter,captionpos=b,label=lst:vc_inverter,language=json]
{
  "@context":[..],
  "id": "did:sov:sg:inverter:vc:123456789",
  "type": ["VerifiableCredential","InverterVC"],
  "issuer": "did:sov:sg:manufacturer:123456789",
  "issuanceDate": "2023-04-01T10:11:12Z",
  "credentialSubject": {
    "immutable": {
      "id": "did:sov:sg:inverter:123456789",
      "serialNo": "123456789",
      "manufacturedDate":"2021-01-01T00:01:02Z",
      ..
    },
    "updatable": {
        "owner": "did:sov:sg:user:123456789",
        "status" : "active",
        "softwareVersion": "v2.0",
        "timelyUpdated":true,
        "missingUpdates":false
        ..
    },
    "firmwareHistory": {
        "v2.0": "2023-04-01T08:11:12Z",
        "v1.8": "2022-11-17T16:34:20Z",
        "v1.0": "2022-01-29T02:56:43Z"
    },
    "resetHistory":{
        |\color{blue} //factory reset date-times|
    }
  },
  ...
  "proof": {..}
}
\end{lstlisting}

The manufacturer then validates the update-specific value to ensure the installation attributes (version, date, and time). Upon successful validation, the manufacturer generates the VC ($VC_{M_i}^{SI_i}$) that reflects the changes in the update history once the ownership is verified. It also invalidates any previously issued VC for the inverter. The new VC changes the current `softwareVersion' attribute in the VC and appends the version and the datetime to the `firmwareHistory' section accordingly. A sample inverter VC is depicted in Listing \ref{lst:vc_inverter}. The manufacturer signs the VC and sends it to the inverter. The inverter stores the VC in its wallet once it has verified that the issuer is the manufacturer. The detailed communication protocol and interactions among entities are depicted in Algorithm 1.

The manufacturer decrypts the message using its private key and then proceeds to validate the authenticity of the message using the public key listed in the user's DIDdoc. Furthermore, it verifies the ownership of the inverter by referring to the `controller' property stated in the inverter's DIDdoc or the current VC issued for the user, thereby confirming ownership. Failure to satisfy either of these conditions results in the manufacturer rejecting the issuance of a new VC.

The next step involves the manufacturer validating the update-specific values to ensure the version, installation date, and time. Upon successful validation, the manufacturer generates a new VC ($VC_{M_i}^{SI_i}$) to reflect the changes in the update history. The manufacturer also invalidates any previously issued VC for the inverter. The new VC includes modifications to the current `softwareVersion' attribute in the VC and appends the version and datetime details to the `firmwareHistory' section as required. A sample inverter VC is shown in Listing \ref{lst:vc_inverter}. Subsequently, the manufacturer signs the VC, encrypts the signed VC using the user/inverter's public key, and transmits it to the inverter/user. The user/inverter decrypts the message and ensures the authenticity of the message. The new VC will be stored in the inverter's wallet after verifying the issuer's authenticity. Additionally, previously issued VCs must be removed from the wallet. Detailed communication protocols and interactions among entities are depicted in Algorithm 1.

\begin{table}
\label{algo:firmware_update}
\begin{tabular}{p{0.14\linewidth}  p{0.75\linewidth}}
\toprule
\multicolumn{2}{l}{\textbf{Algorithm 1:} Firmware Update}\tabularnewline 
\midrule
$M$ & \begin{tabular}{@{}l@{}}
Generate a firmware version ($FW_{M_i}^i$)\\
Publish the firmware to the update server/cloud\\
Generate a new VC ($VC_{M_i}^{FW_i}$) for the firmware
\end{tabular}\tabularnewline\hline

$M \rightarrow SC$ & \begin{tabular}{@{}l@{}}
Send the new update request\\
\quad $msg=VC_{M_i}^{FW_i}||DID^P_{M_i} || nonce_1$\\
\quad $sig=sign(hash(msg),SK^{DID}_{M_i})$ \\
\quad $newUpdReq=(msg || sig)$
\end{tabular}\tabularnewline\hline

$SC$ & \begin{tabular}{@{}l@{}}
Verify the signature and request using $PK^{DID}_{M_i}$\\
Store the $VC_{M_i}^{FW_i}$ on blockchain/db/ipfs\\
Fire update available event
\end{tabular}\tabularnewline\hline

$SI$ & \begin{tabular}{@{}l@{}}
Listen to update events\\
Receive update event
\end{tabular}\tabularnewline\hline 

$SI \rightarrow M$ & Request update ($FW_{M_i}^i$)\tabularnewline\hline

$M \rightarrow SI$ & Send update ($FW_{M_i}^i$)\tabularnewline\hline

$SI$ & \begin{tabular}{@{}l@{}} 
Obtain $VC_{M_i}^{FW_i}$ from DID/VC-API/Wallet\\
Obtain $DIDDoc_{M_i}^{P}$ from DID/VC-API\\
Compare hash values\\
\quad $H_{FW_i}^{authentic}\leftarrow \mbox{Hash value stated on }VC_{M_i}^{FW_i}$\\
\quad $H_{FW_i}^{received}\leftarrow hash(FW_{M_i}^i)$\\
\quad $H_{FW_i}^{authentic}==H_{FW_{M_i}^i}^{received}$\\
Verify signature \\
\quad $sig \leftarrow signature(VC_{M_i}^{FW_{M_i}^i})$\\
\quad $verify(sig, hash(VC_{M_i}^{FW_{M_i}^i}), PK(DIDDoc_{M_i}^{P}))$\\
Install $FW_{M_i}^i$\\
\quad $proof\_value (P_i)\leftarrow$ \\ \quad \quad \quad $firmwareVerifier(FW_{M_i}^i) || timestamp (t_i)$
\end{tabular}\tabularnewline\hline

$U\rightarrow M$ & \begin{tabular}{@{}l@{}} 
Send a new VC request\\
\quad $msg=DID_{{SI_i}{M_i}}^P || DID_{U_i}^{P} || P_i || VC_{M_i}^{SI_i} || t_j || nonce_2 $\\
\quad $sig=sign(hash(msg), SK_{U_i}^{DID})$\\
\quad $newVCReq=(msg || sig)$\\
\quad $enryptVCReq=encrypt(newVCReq, PK^{DID}_{M_i})$
\end{tabular}
\tabularnewline\hline

$M$ & \begin{tabular}{@{}l@{}} 
Decrypt the message using $SK^{DID}_{M_i}$\\
Verify the request using $PK^{DID}_{U_i}$\\ 
Generate a new VC ($VC_{M_i}^{SI_i}$) for the inverter\\
Invalidate old VCs
\end{tabular}\tabularnewline\hline

$M \rightarrow U$ &  \begin{tabular}{@{}l@{}} 
Send the $VC_{M_i}^{SI_i}$\\
\quad $msg=VC_{M_i}^{SI_i} || (nonce_2+1) $\\
\quad $sig=sign(hash(msg), PK_{M_i}^{DID})$\\
\quad $newVCRes=(msg || sig)$\\
\quad $enryptVCRes=encrypt(newVCRes, PK^{DID}_{U_i})$
\end{tabular}\tabularnewline\hline

$U$ & \begin{tabular}{@{}l@{}} 
Decrypt the message using $SK^{DID}_{U_i}$\\
Verify the request using $PK^{DID}_{M_i}$
\end{tabular}\tabularnewline\hline

$U \rightarrow SI$ & Upload $VC_{M_i}^{SI_i}$ to the inverter\tabularnewline\hline

$SI$ & \begin{tabular}{@{}l@{}}Check if the VC's issuer is manufacturer\\
\quad $issuer(VC_{M_i}^{SI_i}) == DID_{M_i}^P$\\
Store the $VC_{M_i}^{SI_i}$\\
Remove old VCs
\end{tabular}\tabularnewline

\bottomrule
\end{tabular}
\end{table}

\subsection{Part 2: Secure VPP Operation}
Given the existence of a VPP, some consumers enrol their inverters in the VPP. These enrolments allow the VPP operator to read the status and send control signals to those inverters. Nevertheless, the VPP operator has no control or visibility over the firmware updates of the enrolled inverters, which makes the entire system vulnerable. Compromised inverters can provide inaccurate information (data/status) and potentially act as an entry point to the system. Addressing this vulnerability, we propose categorising inverters into different trust levels based on their firmware update history. Each trust level should have a well-defined set of interactions, with reduced interactions for lower trust levels. However, it is important to note that the trust state should not be a global attribute of the inverter itself, as different use cases may require different levels of trust. Instead, the trust state in the VPP context is defined by the VPP operator and only applicable within the specific VPP operator's operations.

\begin{figure}
    \centering
    \includegraphics[width=\linewidth]{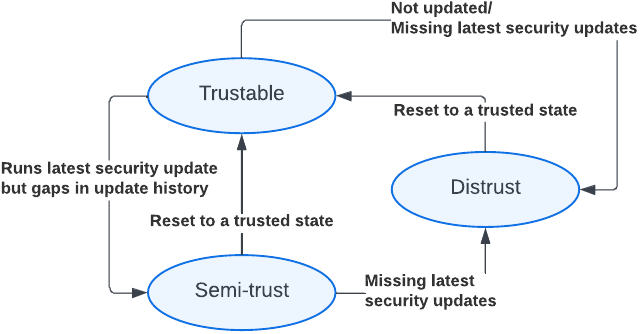}
    \caption{Proposed trust cycle for inverter based on update history}
    \label{fig:trust_cycle}
\end{figure}

Figure \ref{fig:trust_cycle} depicts the proposed trust cycle for inverters. We define three trust levels: Trustable, Semi-Trust, and Distrust. The trust state can change throughout the inverter's lifecycle contemplating its firmware update history. An inverter is initially assumed to be in the `Trustable' state when it leaves the manufacturer. It remains in the `Trustable' state if it updates the firmware as soon as possible or within a reasonable period upon the availability of a new update. The VPP operator can define this reasonable period based on the severity of the security vulnerabilities in the previous version. If an inverter fails to update its firmware within a reasonable period, the trust state changes to `Semi-Trust'. An inverter can be in the `Semi-Trust' state even if it operates the latest version of its firmware, indicating a delay in updating to the newest version. Further, an inverter that runs (or pretends to run) the latest security update with significant gaps (time, versions) in the update history is considered a `Semi-Trust' device. Inverters in `Semi-Trust' possess an increased risk of being compromised and can potentially mimic its capabilities, configurations and status. 

An inverter's trust state changes to `Distrust' if it is not updated or not running the latest security-related firmware. Additionally, given the availability of associated CVEs of a firmware version, the VPP operator can categorise inverters as `distrust' if the manufacturer fails to provide updates fixing the security issues within a reasonable time window. An inverter in the `Distrust' or `Semi-Trust' state can only be set back to `Trustable' if it undergoes a factory reset and is installed with all the updates. The manufacturer should embed a method to determine the factory reset date and time. Similar to when installing a firmware update, the user can request a new VC for the inverter. The manufacturer issues a new VC (and invalidate current VC) to reflect the factory reset history. The updated VC includes the reset date and time in the `resetHistory' section in the VC (see Listing \ref{lst:vc_inverter} for an example).

The proposed trust cycle can be implemented at the VPP operator, leveraging the `timelyUpdated' and `missingUpdates' attributes stated in the inverter VC. In this case, the manufacturer decides the time threshold to determine if an inverter is updated within a reasonable time window. However, the VPP operator may need to tighten or relax the time window (threshold) based on their use case and specific vulnerabilities. Thus, the operators should be able to possess their implementation of the `timelyUpdated' attribute. The VPP operator can use the inverter's VCs and the list/VC of the available updates issued by the manufacturer. The manufacturer can provide published updates for a specific inverter model. A simple REST API can be leveraged to provide a list of updates for a given inverter model. The VPP operator can verify a specific inverter's update history (on the inverter's VC) against the available update list. If the inverter possesses the latest update and all the updates are installed within the predefined reasonable time window, the inverter is considered to be in the `trustable' state. The inverter's trust state is `semi-trust' or `distrust' if there are delays in installing updates or missing updates, as described. Algorithm \ref{algo:trsut_state} depicts the pseudocode to determine the trust state of an inverter.

\setcounter{algocf}{1}
\begin{algorithm}
\SetAlgoLined 
\KwIn{inverter's VC ($VC^{SI_i}_{M_i}$), available update list and threshold $T$} 
\KwOut{Trust state for the inverter} 
\BlankLine
$trustState \gets $``distrust"\\
$timelyUpdated \gets $True\\
$allUpdates \gets $True\\
\BlankLine
\If{$resetHistory$ exist in $VC^{SI_i}_{M_i}$}{
    $latestResetTime \gets top(VC^{SI_i}_{M_i}[``resetHistory"])$ \\
    modify the updateList to only reflect updates published after the latestResetTime
}
\BlankLine
\For{each availableUpdate($AvUpdt_i$), availableTime ($AvUpdt_i^t$) in the updateList}{
    $missedUpdate \gets $True\\
    \For{each installUpdate($InstlUpdt_i$), installTime ($InstlUpdt_i^t$) in the $VC^{SI_i}_{M_i}$}{
        \If{$AvUpdt_i==InstlUpdt_i$}{
            $missedUpdate \gets $False\\
            $timelyUpdated \gets timelyUpdated\,$AND$\,(AvUpdt_i^t + $T \textgreater $InstlUpdt_i^t)$\\
            break
        }
    }
    \BlankLine
    $allUpdates \gets allUpdates\,$AND$\,!missedUpdate$\\
    \BlankLine
    \If{missedUpdate == True}{
        break
    }
}
\If{$allUpdates$}{
    \eIf{$timelyUpdated$}{
        $trustState \gets $``trustable"\\
    }
    {
        $trustState \gets$ ``semi-trust"\\
    }
}
\caption{Trust state calculation pseudo code}
\label{algo:trsut_state}
\end{algorithm}

The VPP operator can fully utilise the data from the inverters in the `trustable' state in all operations/optimisations/calculations and issue authorised control signals. Data from the inverters in the `semi-trust' state can be consumed with some verification or uncertainty factors. However, the operator cannot ensure the reliable execution of the control signals by these inverters. Operators should not use data from inverters in the `distrust' state for any operation, as the inverters in that state have a significant probability of being compromised. 

\section{Comparison with IEEE 2030.5 and CSIP}
IEEE 2030.5~\cite{ieee2030} and the Common Smart Inverter Profile (CSIP)~\cite{smart_common_2022} are widely used and commonly adopted application protocols/standards and implementation guidelines in the smart inverter context globally. Table \ref{tab:comparison} compares existing standards with the proposed solution. It should be noted that while existing standards cover the entire inverter life cycle, our solution focuses solely on analysing the firmware update process. However, we have considered the potential to extend our proposed solution in this comparison.

CSIP and IEEE 2030.5 standards utilise the fingerprint of manufacturer-issued device certificates to derive inverter identities (Long-form Device Identifier and Short-form Device Identifier). This methodology produces a static identity and assumes that the device certificate is valid indefinitely. However, real-world scenarios often require the revocation or renewal of initially issued certificates and keys. Thus, maintaining the longevity of inverter identity becomes complex, as it is closely tied to the device certificate. In contrast, our approach assumes a manufacturer assigned DID, eliminating the tight coupling with a specific key. This approach allows for key pair revocation and renewal without altering the identity. Further, the use of a one-way hash chain can address this limitation if the DID is generated based on a private/public key~\cite{did_onewayhash}.

Leveraging DIDs makes key management (rotation, revocation, renewal) standardised and convenient, whereas the existing standards do not specify the key management process. DIDs use distributed trust in contrast to the traditional centralised PKI-based root-of-trust approach. This distributed trust enables verification for inverters, even if the manufacturer's existence is compromised in the future, considering the long lifespan of inverters.
 
The IEEE 2030.5 standard specifies 12 device attributes, including location, identity, firmware version, and hardware version, under the `DeviceInformation' package. These attributes help to construct a more reliable and precise assessment of inverters, as utilities and VPP operators do not have complete control or visibility over inverters. Our design extends this further by allowing manufacturers to use VCs in sharing additional information with appropriate security mechanisms, ensuring that authorised entities can access only the necessary information. Similarly, manufacturers can utilise the VC-based approach to specify additional capabilities, functionalities, and limits of their inverters other than those that are already covered under the IEEE standard. Further, regulators can define VC schemes to standardise the additional information and capabilities instead of duplicating schemes.

Despite the current availability of attributes and capabilities, existing standards and guidelines lack a mechanism to track changes to these attributes and capabilities over time. Furthermore, adversaries can fabricate inaccurate information, such as location, capabilities, and firmware version, especially after a firmware/software compromise, as the standards do not encompass any post-compromise aspects. Although the proposed model does not eliminate attacks, it offers a mechanism for stakeholders to assess the risk associated with compromised inverters. This enhanced assessment capability empowers stakeholders to adjust their interactions with these inverters. Moreover, our approach has the potential for extension to incorporate the historical changes in attributes and capabilities. Much like the history of firmware updates, having access to this information can facilitate comprehensive anomaly detection and risk assessment for operators.

\begin{table*}
\centering
\caption{Comparison with existing standards}
\label{tab:comparison}
\begin{tabular}{lll}
\toprule
Criteria & IEEE 2030.5/CSIP & Ours\tabularnewline
\midrule
Identity & Derived from the certificate fingerprint & DID\tabularnewline
Validity & Indefinitely & Customisable\tabularnewline
Certificate Revocation & Not specified & Supported via DIDdocs\tabularnewline
Root of Trust & PKI & Distributed\tabularnewline
Device information & Limited (Location, ID, Hardware version, Model No, etc.) & Unlimited\tabularnewline
Device capabilities & Supported & Supported\tabularnewline
History of attributes & Not supported & Supported via VC\tabularnewline
Format for device information & Not specified & VC scheme\tabularnewline
Format for capabilities & Not specified & VC scheme\tabularnewline
\bottomrule
\end{tabular}
\end{table*}

\section{Evaluation}

\subsection{Security Analysis}
\subsubsection{Security Properties}
This section provides semi-formal proof to guarantee the security requirements defined in the System model section. 

\paragraph{Confidentiality}
Private information associated with a particular inverter, including the firmware version and the update history, is stored in the manufacturer-issued VC. This VC is securely stored in a digital wallet inside the inverter (or the owner's wallet), which is assumed to be secure. Consequently, as long as an adversary does not compromise the digital wallet, the content within the VC remains confidential. To access the VC during transit, the adversary needs to compromise the encryption or session key. We assume the presence of secure channels between the manufacturer and the inverter and between the VPP operator and the inverter, which utilises existing well-established transport layer security protocols. Therefore, confidentiality is maintained unless the associated keys or the digital wallet are compromised.

\paragraph{Integrity}
The inverter VC contains all the information about the inverter that other stakeholders may require. This VC is issued by the inverter manufacturer and includes at least one proof element, typically a digital signature, to ensure the authenticity and integrity of the VC's content. Although adversaries may attempt to mimic incorrect information about an inverter, they cannot replicate the manufacturer-issued VC without access to the manufacturer's private key. This private key is also associated with the public key declared on the manufacturer's DIDdoc for verification. Similarly, adversaries cannot deceive inverters with fake updates, as the `Update VC' has tamper-evident properties. Consequently, the integrity of the system is maintained unless the manufacturer's private key is compromised or the manufacturer fails to include at least one proof element in the VC.

\paragraph{Availability}
We assume that the inverter/owner requires sharing the inverter VC on request. The associated public key used to sign VCs is linked to the manufacturer's DIDdoc and is anchored on the blockchain, which is designed to be continuously available. Therefore, the blockchain ensures the continuous availability of the manufacturer's DIDdoc and, consequently, the inverter VCs upon request. This availability ensures that anyone, at any time, can verify the validity of the presented VCs, demonstrating the system's availability for verification purposes.

\paragraph{Authentication}
The use of digital signatures ensures the authentication property in the proposed system. More specifically, firmware updates and VCs (update-related and inverter) consist of the manufacturer's signature. New VC requests are signed by the inverter owner. Therefore, the proposed system can guarantee the authentication property for the desired processes unless the associated private keys are compromised.

\paragraph{Authorisation}
We assume the existence of a reliable method for installing updates on inverters, whether initiated by the inverter itself, the owner, or some other authorised entity. Therefore, we have focused our analysis on the authorisation property, particularly on requesting a new VC for the installed updates and verifying inverter information.

Requesting a new VC for the inverter necessitates the provision of the current VC of the inverter, the owner's DID, and other required information. Furthermore, the request must be digitally signed by the owner. Given our assumption that only the legitimate owner of the inverter can request a new VC, adversaries would need to forge the owner's digital signature to imitate authorisation. Additionally, the requester would have to obtain the current VC of the inverter, which is securely stored in the digital wallet. Thus, the proposed system ensures proper authorisation before issuing new VCs for inverters, as long as the owner's private key and the security of the digital wallet remain uncompromised. Similarly, regarding the verification aspect, when the owner/inverter presents the inverter's VC to the verifier with consent, it implies possession of the inverter's VC and the associated keys necessary to establish proper authorisation. 

\paragraph{Non-repudiation}
The manufacturer is responsible for issuing VCs for all firmware updates they release. These VCs are signed by the manufacturer and stored on the blockchain through a smart contract. Smart contracts are immutable once they are deployed, and only the manufacturer has the authority to update the blockchain via the contract. Consequently, the manufacturer cannot deny publishing VCs associated with the firmware updates they release. Furthermore, the smart contract deployer can be traced by examining the transaction history, which ensures that the manufacturer cannot disclaim the deployment of the smart contract responsible for managing VCs related to updates. Thus, the proposed system guarantees non-repudiation regarding released firmware updates, unless the manufacturer's private key is compromised.

From an inverter's perspective, it is assumed that the manufacturer has a mechanism to determine the firmware version and the installation date and time. The system requires the inverters to present update-specific values to ensure that they have successfully installed a particular update. Additionally, we assume that the firmware does not permit downgrading to an earlier version. These combined properties ensure that inverters cannot deny the installation of a specific firmware version after it has been installed.

\subsubsection{Resiliency against Attacks}
\paragraph{Mimic incorrect meta information of an inverter}
Users are required to provide the inverter's VC to the VPP operator when registering their inverter. This VC is issued by the manufacturer, and the meta-information (capabilities of the inverter) is verified by the manufacturer before issuing the VC. Thus, adversaries cannot provide a new VC with a different set of capabilities as the manufacturer must issue it. Therefore, the proposed system is robust against the falsification of an inverter's capabilities, unless the manufacturer's private key is compromised where the adversary can issue a fake VC using the compromised manufacturer's key.

\paragraph{Inverter with an outdated firmware version}
The inverter enrollment request includes the inverter VC issued by the manufacturer, which accurately reflects the current firmware version. The adversary cannot mimic the firmware version, as the VC is issued using a challenge-response mechanism that verifies the installed firmware version. Thus, the proposed design is robust against enrolling inverters with outdated firmware versions. However, an advanced adversary may attempt to trick the system by downgrading the firmware after successfully obtaining the VC for the current firmware. The firmware can prevent this attack by not allowing firmware downgrades.

\paragraph{Inverter reside outside the geographical area}
Our model does not provide a direct solution for this. However, the model can be modified to include the geolocation of the user or inverter in the inverter VC. Having geolocation information in the inverter VC can ensure that only the inverters located within a specific geographical area are considered for VPP operations. Nevertheless, this would require an authorised entity, such as manufacturers, authorised sellers, or installers, to verify the installed location during the inverter installation process.

\subsection{Performance}
Table \ref{tab:performance} presents key performance metrics, including the average time for DIDdoc generation, size of update history VC, VC generation time, and VC verification time. The implementation utilised Java, ECDSA-based key-pairs, and JSON Web Token (JWT) representations for the VCs. This evaluation involved the generation of 100,000 VCs and DIDdoc for rigorous assessment.

\begin{table}[]
\caption{Performance evaluation}
\centering
\begin{tabular}{llll}
\toprule
DIDdoc gen time & VC size & VC gen time & VC verify time\\
\midrule
0.0007 ms & 2.3 kB  & 1.16 ms & 1.04 ms \tabularnewline
\bottomrule
\end{tabular}
\label{tab:performance}
\end{table}

The proposed framework offers the necessary scalability and minimal overhead, making it highly suitable for managing firmware updates in the VPP context. Its lightweight design ensures minimal impact on grid operations, preserving overall performance and responsiveness. The overhead of DIDdoc generation is 0.0007 milliseconds (ms) per inverter for manufacturers. However, it is important to note that key-generation and the actual storing of DIDdocs might introduce additional overhead.
The JSON representation of the update history VC (Listing \ref{lst:vc_inverter}) averages 1.3 kB in size and 2.3 kB with the signature, effectively minimising both storage and transmission overhead. Different types of cryptographic keys and optimisation techniques can further reduce the VC size. Moreover, the VC generation and verification times are 0.017 ms and 1.04 ms, respectively. It facilitate real-time trust establishment and decision-making. These attributes are essential for ensuring the resilience and security of VPPs, emphasising the practicality and effectiveness of the proposed framework in smarty grid operations.

\section{Discussion}
In this work, we assumed the existence of a secure channel between the manufacturer and the inverter for both communication and the update delivery process. Existing encryption-based secure communication mechanisms can be adopted for this purpose. However, we acknowledge that manufacturers can also implement a proprietary communication method, leveraging the availability of public key information in the DIDdoc. The firmware installation process is independent of the use of VC. Therefore, the proposed framework supports offline updates. Offline firmware updates involve the user downloading the update and installing it on the inverter upon successfully verifying the hash. Manufacturers can delegate the firmware verification process to the firmware itself to prevent errors or mistakes that may occur during human-based verification or unauthorised attempts. If the update is conducted offline, the user or inverter can request the VC (to reflect the new version) using the proof value obtained from the update.

A public or consortium blockchain can be utilised in implementing the proposed solution. However, a permissioned blockchain (managed by manufacturers and VPP operators) is the better option, given the limited number of stakeholders and the unnecessary need for the public availability of the information. Moreover, there is no strict dependency between the use of DIDs and the smart contracts that manage firmware and updates. Therefore, it is possible to create a separate blockchain to manage the DIDs (via the DID API). However, it is essential to ensure cross-chain communication, as smart contracts require manufacturers' identity. Existing DID and wallet implementations can be utilised in implementing the proposed framework instead of developing novel methods. Inverter owners should be able to request VCs from the manufacturer on behalf of the inverters, where the owner's identity can be linked from the same blockchain or other mechanisms. The firmware update process can be further secured by integrating an antivirus scan for the downloaded firmware \cite{Ahn2021}. 

Despite the convenience of utilising third-party dependencies in developing the framework, dependencies can introduce additional vulnerabilities. Furthermore, manufacturers may need to provide additional updates to address security vulnerabilities in the dependencies, which can be an overhead. Therefore, it is essential to limit the use of third-party libraries and frameworks and only employ highly stable, well-tested ones to enhance the security of the updates and minimise the overhead of releasing additional firmware updates.

Publishing update VCs with CVEs can potentially raise concerns, as adversaries may learn about vulnerabilities in previous versions. Omitting CVE numbers from the update VC can address this issue. However, we believe that the association between the firmware version and the addressed CVE numbers should be accessible to authorised entities. Therefore, we propose introducing a separate VC to capture the associated CVE numbers, which would not be publicly accessible. Instead, it would only be shared with authorised stakeholders, such as DNSPs and VPP operators, on request. A more robust approach is to implement a zero-knowledge proof-based system that exposes an interface where interested stakeholders can interact to confirm whether a particular firmware addresses specific CVEs.

The VPP operator can implement more granular levels of the trust cycle depending on their level of interaction. Additionally, a phased approach could be designed to gradually restrict interactions with inverters based on their trust levels, further enhancing the security and reliability of the system. 

Large-scale implementation of advanced smart grid concepts, such as VPPs, aggregators, and energy trading platforms, is inevitable with the increased use of DERs. Consequently, the deployment of smart inverters for managing DERs is apparent. Thus, ensuring the security of these smart inverters is essential to prevent compromises that can lead to severe consequences, including loss of lives and substantial financial damage in nation states' critical infrastructure operations.

\section{Conclusion \& Future Work}
This work used VCs to ensure the availability of security-relevant information on firmware updates for smart inverters. We also detailed the restructuring of the firmware update delivery process using VCs. Furthermore, we demonstrated how to leverage the accessible security metadata to categorise smart inverters into different trust levels. External entities engaging with these smart inverters gain a comprehensive view of the inverters with the availability of security metadata, including firmware versions and update history. This approach can enhance the implementation of more reliable and secure smart grid solutions.

Future work should focus on performing analyses of various alternatives, such as different DID methods, VC flavours, wallets, and blockchains, to identify the most efficient technology stack to implement this framework in practice. Also, formal security proofs must be derived to guarantee the security properties of the proposed framework. 

\bibliographystyle{IEEEtran}
\bibliography{references}

\begin{thebibliography}{10}
\providecommand{\url}[1]{#1}
\csname url@samestyle\endcsname
\providecommand{\newblock}{\relax}
\providecommand{\bibinfo}[2]{#2}
\providecommand{\BIBentrySTDinterwordspacing}{\spaceskip=0pt\relax}
\providecommand{\BIBentryALTinterwordstretchfactor}{4}
\providecommand{\BIBentryALTinterwordspacing}{\spaceskip=\fontdimen2\font plus
\BIBentryALTinterwordstretchfactor\fontdimen3\font minus
  \fontdimen4\font\relax}
\providecommand{\BIBforeignlanguage}[2]{{%
\expandafter\ifx\csname l@#1\endcsname\relax
\typeout{** WARNING: IEEEtran.bst: No hyphenation pattern has been}%
\typeout{** loaded for the language `#1'. Using the pattern for}%
\typeout{** the default language instead.}%
\else
\language=\csname l@#1\endcsname
\fi
#2}}
\providecommand{\BIBdecl}{\relax}
\BIBdecl

\bibitem{solarenergyAu}
``Solar energy,'' \url{https://arena.gov.au/renewable-energy/solar/}, accessed:
  2023-04-14.

\bibitem{solarenergyEU}
``Solar energy,''
  \url{https://energy.ec.europa.eu/topics/renewable-energy/solar-energy\_en},
  accessed: 2023-04-14.

\bibitem{saboori2011virtual}
H.~Saboori, M.~Mohammadi, and R.~Taghe, ``Virtual power plant (vpp),
  definition, concept, components and types,'' in \emph{2011 Asia-Pacific power
  and energy engineering conference}.\hskip 1em plus 0.5em minus 0.4em\relax
  IEEE, 2011, pp. 1--4.

\bibitem{aemoinverters2020}
\BIBentryALTinterwordspacing
``{Renewable Integration Study Stage 1 Appendix A: High Penetrations of
  Distributed Solar PV Important notice PURPOSE},'' Australian Energy Market
  Operator, Tech. Rep., 2020. [Online]. Available:
  \url{https://www.aemo.com.au/energy-systems/Major-publications/Renewable-Integration-Study-RIS.}
\BIBentrySTDinterwordspacing

\bibitem{qi2016cybersecurity}
J.~Qi, A.~Hahn, X.~Lu, J.~Wang, and C.-C. Liu, ``Cybersecurity for distributed
  energy resources and smart inverters,'' \emph{IET Cyber-Physical Systems:
  Theory \& Applications}, vol.~1, no.~1, pp. 28--39, 2016.

\bibitem{johnson2017design}
J.~Johnson, J.~Flicker, A.~Castillo, C.~Hansen, M.~El-Khatib, D.~Schoenwald,
  M.~Smith, R.~Graves, J.~Henry, T.~Hutchins \emph{et~al.}, ``Design and
  implementation of a secure virtual power plant,'' \emph{no. September}, pp.
  243--287, 2017.

\bibitem{Alagappan2022}
\BIBentryALTinterwordspacing
A.~Alagappan, S.~K. Venkatachary, and L.~J.~B. Andrews, ``Augmenting zero trust
  network architecture to enhance security in virtual power plants,''
  \emph{Energy Reports}, vol.~8, pp. 1309--1320, 2022. [Online]. Available:
  \url{https://doi.org/10.1016/j.egyr.2021.11.272}
\BIBentrySTDinterwordspacing

\bibitem{zhang2018security}
F.~Zhang and Q.~Li, ``Security vulnerability and patch management in electric
  utilities: A data-driven analysis,'' in \emph{Proceedings of the First
  Workshop on Radical and Experiential Security}, 2018, pp. 65--68.

\bibitem{Choi2021}
J.~Choi, B.~Ahn, S.~Pedavalli, S.~Ahmad, A.~Villasenor, and T.~Kim, ``Secure
  firmware update and device authentication for smart inverters using
  blockchain and phyiscally uncloable function (puf)-embedded security
  module.''\hskip 1em plus 0.5em minus 0.4em\relax Institute of Electrical and
  Electronics Engineers Inc., 2021.

\bibitem{konstantinou2015impact}
C.~Konstantinou and M.~Maniatakos, ``Impact of firmware modification attacks on
  power systems field devices,'' in \emph{2015 IEEE International Conference on
  Smart Grid Communications (SmartGridComm)}.\hskip 1em plus 0.5em minus
  0.4em\relax IEEE, 2015, pp. 283--288.

\bibitem{Bere2021}
G.~Bere, B.~Ahn, J.~J. Ochoa, T.~Kim, A.~A. Hadi, and J.~Choi,
  ``Blockchain-based firmware security check and recovery for smart
  inverters.''\hskip 1em plus 0.5em minus 0.4em\relax Institute of Electrical
  and Electronics Engineers Inc., 6 2021, pp. 675--679.

\bibitem{Ahn2021}
B.~Ahn, G.~Bere, S.~Ahmad, J.~Choi, T.~Kim, and S.~W. Park,
  ``Blockchain-enabled security module for transforming conventional inverters
  toward firmware security-enhanced smart inverters.''\hskip 1em plus 0.5em
  minus 0.4em\relax Institute of Electrical and Electronics Engineers Inc.,
  2021, pp. 1307--1312.

\bibitem{Ansay2019}
R.~Ansay, J.~Kempf, O.~Berzin, C.~Xi, and I.~Sheikh, ``Gnomon: Decentralized
  identifiers for securing 5g iot device registration and software
  update.''\hskip 1em plus 0.5em minus 0.4em\relax IEEE, 2019, pp. 1--6.

\bibitem{didw3c}
\BIBentryALTinterwordspacing
M.~Sporny, D.~Longley, M.~Sabadello, D.~Reed, O.~Steele, and C.~Allen,
  ``Decentralized identifiers (dids) v1.0,'' W3C, Tech. Rep., 2021. [Online].
  Available: \url{https://www.w3.org/TR/did-core/}
\BIBentrySTDinterwordspacing

\bibitem{vcw3c}
\BIBentryALTinterwordspacing
M.~Sporny, D.~Longley, and D.~Chadwick, ``{Verifiable Credentials Data Model
  1.1},'' W3C, Tech. Rep., 2022. [Online]. Available:
  \url{https://www.w3.org/TR/vc-data-model/}
\BIBentrySTDinterwordspacing

\bibitem{totp}
D.~M'Raihi, S.~Machani, M.~Pei, and J.~Rydell, ``Totp: Time-based one-time
  password algorithm,'' Tech. Rep., 2011.

\bibitem{ieee2030}
``Ieee standard for smart energy profile application protocol,'' \emph{IEEE Std
  2030.5-2018 (Revision of IEEE Std 2030.5-2013)}, pp. 1--361, 2018.

\bibitem{smart_common_2022}
C.~Smart, I.~Profile, and W.~Group, ``Common {Smart} {Inverter} {Profile} -
  {Australia},'' Tech. Rep., 2022, issue: March.

\bibitem{did_onewayhash}
C.-S. Park and H.-M. Nam, ``A new approach to constructing decentralized
  identifier for secure and flexible key rotation,'' \emph{IEEE Internet of
  Things Journal}, vol.~9, no.~13, pp. 10\,610--10\,624, 2022.

\end{thebibliography}

\end{document}